\documentclass[useAMS,usenatbib]{mn2e}

\usepackage{graphicx}

\newcommand{\ion}[2]{#1\,{\sc #2}}

\title[Multiwavelength study of the Of?p star CPD\,$-$28$^{\circ}$\,2561]{
New multiwavelength observations of the Of?p star CPD\,$-$28$^{\circ}$\,2561
}

\author[Hubrig et al.]{
S.~Hubrig$^{1}$\thanks{E-mail: shubrig@aip.de},
M.~Sch\"oller$^{2}$,
A.~Kholtygin$^{3}$,
H.~Tsumura$^{4}$,
A.~Hoshino$^{4}$,
\and
S.~Kitamoto$^{4}$,
L.~Oskinova$^{5}$,
R.~Ignace$^{6}$,
H.~Todt$^{5}$,
I.~Ilyin$^{1}$\\
$^1$ Leibniz-Institut f\"ur Astrophysik Potsdam (AIP), An der Sternwarte 16, 14482 Potsdam, Germany \\
$^2$ European Southern Observatory, Karl-Schwarzschild-Str.~2, 85748 Garching bei M\"unchen, Germany \\
$^3$ Astronomical Institute, St.~Petersburg State University, Universitetski pr.~28, 198504, St.~Petersburg, Russia \\
$^4$ Department of Physics, College of Science,
Rikkyo University 3-34-1, Nishi-Ikebukuro, Toshima--ku, Tokyo 171--8501, Japan \\
$^5$ Universit\"at Potsdam, Institut f\"ur Physik und Astronomie, 14476~Potsdam, Germany \\
$^6$ Department of Physics \& Astronomy, East Tennessee State University, Box 70652, Johnson City, TN, 37614, USA
}

\begin{document}

\date{Accepted Received; in original form}

\pagerange{\pageref{firstpage}--\pageref{lastpage}} \pubyear{2014}

\maketitle

\label{firstpage}

\begin{abstract}
A rather strong mean longitudinal magnetic field of the order of a few 
hundred Gauss was detected a few years ago in the Of?p star CPD\,$-28^{\circ}$\,2561 
using FORS2 low-resolution spectropolarimetric observations.
In this work we present additional low-resolution spectropolarimetric observations obtained during 
several weeks in 2013 December using FORS\,2 (FOcal Reducer low dispersion Spectrograph) mounted at the 8-m Antu 
telescope of the VLT. These observations cover a little less than half of the stellar rotation period 
of 73.41\,d mentioned in the literature. The behaviour of the mean longitudinal magnetic field is 
consistent with the assumption of a single-wave variation during the stellar rotation cycle,
indicating a dominant dipolar contribution to the magnetic field topology. The estimated polar strength 
of the surface dipole $B_d$ is larger than 1.15\,kG.
Further, we compared the behaviour of the line profiles of various elements
at different rotation phases associated with different magnetic field strengths. 
The strongest contribution of the emission component
is observed at the phases when the magnetic field shows a negative or positive extremum. 
The comparison of the spectral behaviour of CPD\,$-28^{\circ}$\,2561 with that of another Of?p star, HD\,148937 of
similar spectral type, reveals remarkable differences in the degree of variability between both stars.
Finally, we present new X-ray observations obtained with the {\em Suzaku X-ray Observatory}. 
We report that the star is X-ray bright with  $\log{L_{\rm X}/L_{\rm bol}}\approx -5.7$.
The low resolution X-ray spectra reveal the presence of a plasma heated up to 24\,MK.
We associate the 24\,MK plasma in CPD\,$-28^{\circ}$\,2561 with the presence 
of a kG strong magnetic field capable to confine stellar wind.
\end{abstract}


\section{Introduction}
\label{sect:intro}

\citet{Walborn1973}
introduced the Of?p category for massive O stars displaying recurrent spectral variations in 
certain spectral lines, sharp emission or P\,Cygni profiles in \ion{He}{i} and the Balmer lines, and 
strong \ion{C}{iii} emission lines around 4650\,\AA{}.
After the first discovery of a longitudinal magnetic field of the order 
of 200\,G in the Of?p star HD\,191612 \citep{Donati2006}, the search for a presence of a magnetic field in remaining
other Of?p stars has been undertaken using FORS\,2 spectropolarimetric observations by the 
MAGORI collaboration and using ESPaDOnS and Narval observations by the MiMeS collaboration. 
Only five Galactic Of?p stars are presently known: HD\,108, NGC\,1624-2, CPD\,$-28^{\circ}$\,2561, HD\,148937, and HD\,191612 
\citep{Walborn2010}, and all of them show evidence for the presence of magnetic fields 
\citep{Martins2010,Wade2012a,Hubrig2008,Hubrig2011a,Hubrig2013,Donati2006}.

Among the five known Of?p stars only the star CPD\,$-28^{\circ}$\,2561 has not been studied in detail yet, neither
with spectroscopy/spectropolarimetry nor in X-rays.
Previous observations of the Of?p star CPD\,$-28^{\circ}$\,2561 by \citet{Hubrig2011a} 
revealed a longitudinal magnetic field of the order of a few hundred Gauss at a significance level
of more that 3$\sigma$. 

Due to the relative faintness of CPD\,$-28^{\circ}$\,2561 with $V$=10.1, it was only 
scarcely studied in the past. 
\citet{Levato1988} acquired radial velocities of 35 OB stars with carbon, nitrogen, and oxygen 
anomalies and found variability of a few emission lines with a probable period of 17\,days.
\citet{Walborn2010} mentioned that CPD\,$-28^{\circ}$\,2561 undergoes extreme spectral transformations very similar 
to those of HD\,191612, on a timescale of weeks, inferred from the variable emission intensity of the
\ion{C}{iii} $\lambda\lambda$~4647--4650--4652 triplet.
Also the spectropolarimetric observations of \citet{Hubrig2013} on three consecutive nights in 2011 
revealed strong variations in several hydrogen and helium line profiles. 

According to \citet{Wade2012}, CPD\,$-28^{\circ}$\,2561 is currently monitored by 
Barb\'a et al.\ (in preparation).
\citet{Petit2013} list in their Table\,1 fundamental parameters
and a rotation period of 70\,d for this star.  
The most recent published result on CPD\,$-28^{\circ}$\,2561 reports on a 73.41-day rotation period, probably
deduced from spectroscopic monitoring \citep{Naze2014}.

The presented study is the continuation of the monitoring of the magnetic field of  CPD\,$-28^{\circ}$\,2561 over
its rotation period  with FORS\,1/2 at the VLT.
Using service mode observations,
we succeeded in covering a significant part of the rotation cycle.
In Sect.~\ref{sect:obs}, we give an overview of our spectropolarimetric observations 
and magnetic field measurements with VLT/FORS\,2, followed by the discussion
of the distinct spectral variability revealed in the FORS\,2 spectra. 
Section~\ref{sect:xrays} is devoted to X-ray observations obtained with the 
{\em Suzaku X-ray Observatory}.
Finally, we summarize the results of our observations in Sect.\,\ref{sect:concl}.

\section{Observations and magnetic field measurements}
\label{sect:obs}

Sixteen new spectropolarimetric observations of the Of?p star CPD\,$-28^{\circ}$\,2561 were carried
out from 2013 December 5 to 30 in service mode at the European
Southern Observatory with FORS\,2 mounted on the 8-m Antu telescope of the
VLT. This multi-mode instrument is equipped with
polarization analyzing optics, comprising super-achromatic half-wave and quarter-wave 
phase retarder plates, and a Wollaston prism with a beam divergence of 22$\arcsec$ in 
standard resolution mode.
Polarimetric spectra were obtained with the GRISM~600B and 
the narrowest slit width of 0$\farcs$4 to achieve 
a spectral resolving power of $R\sim2000$. The use of the mosaic detector 
with a  pixel size of 15\,$\mu$m allowed us to cover a
spectral range, from 3250 to 6215\,\AA{}, which includes all hydrogen Balmer lines 
from H$\beta$ to the Balmer jump. 

From the raw FORS\,2 data, the parallel and perpendicular beams
are extracted using a pipeline written in the MIDAS environment
by T.~Szeifert, the very first FORS instrument scientist.
This pipeline reduction by default includes background subtraction.
A unique wavelength calibration frame is used for each night.

A first description of the assessment of the longitudinal magnetic field
measurements using FORS\,1/2 spectropolarimetric observations was presented 
in our previous work (e.g.\ \citealt{Hubrig2004a,Hubrig2004b}, 
and references therein).

To minimize the cross-talk effect,
and to cancel errors from 
different transmission properties of the two polarised beams,
a sequence of subexposures at the retarder
position angles $-45^{\circ}+45^{\circ}$, $+45^{\circ}-45^{\circ}$,
$-45^{\circ}+45^{\circ}$, etc.\ is usually executed during observations. Moreover, the reversal of the quarter wave 
plate compensates for fixed errors in the relative wavelength calibrations of the two
polarised spectra.
The $V/I$ spectrum is calculated using:
\begin{equation}
\frac{V}{I} = \frac{1}{2} \left\{ 
\left( \frac{f^{\rm o} - f^{\rm e}}{f^{\rm o} + f^{\rm e}} \right)_{-45^{\circ}} -
\left( \frac{f^{\rm o} - f^{\rm e}}{f^{\rm o} + f^{\rm e}} \right)_{+45^{\circ}} \right\}
\end{equation}
where $+45^{\circ}$ and $-45^{\circ}$ indicate the position angle of the
retarder waveplate and $f^{\rm o}$ and $f^{\rm e}$ are the ordinary and
extraordinary beams, respectively.  Rectification of the $V/I$ spectra was
performed in the way described by \citet{Hubrig2014}.
Null profiles, $N$, are calculated as pairwise differences from all available 
$V$ profiles.  From these, 3$\sigma$-outliers are identified and used to clip 
the $V$ profiles.  This removes spurious signals, which mostly come from cosmic
rays, and also reduces the noise. A full description of the updated data 
reduction and analysis will be presented in a separate paper (Sch\"oller et 
al., in preparation).

The mean longitudinal magnetic field, $\left< B_{\rm z}\right>$, is 
measured on the rectified and clipped spectra based on the relation
\begin{eqnarray} 
\frac{V}{I} = -\frac{g_{\rm eff}\, e \,\lambda^2}{4\pi\,m_{\rm e}\,c^2}\,
\frac{1}{I}\,\frac{{\rm d}I}{{\rm d}\lambda} \left<B_{\rm z}\right>\, ,
\label{eqn:vi}
\end{eqnarray} 

\noindent 
where $V$ is the Stokes parameter that measures the circular polarization, $I$
is the intensity in the unpolarized spectrum, $g_{\rm eff}$ is the effective
Land\'e factor, $e$ is the electron charge, $\lambda$ is the wavelength,
$m_{\rm e}$ is the electron mass, $c$ is the speed of light, 
${{\rm d}I/{\rm d}\lambda}$ is the wavelength derivative of Stokes~$I$, and 
$\left<B_{\rm z}\right>$ is the mean longitudinal (line-of-sight) magnetic field.

The longitudinal magnetic field was measured using the entire spectrum including all available absorption
and emission lines.
The feasibility of longitudinal magnetic field measurements in massive stars 
using low-resolution spectropolarimetric observations was demonstrated by previous studies of O and B-type stars
(e.g., \citealt{Hubrig2006,Hubrig2008,Hubrig2009,Hubrig2011a,Hubrig2011b,Hubrig2013}).

Note that we do not differentiate between 
absorption and emission lines, since the relation between the Stokes $V$ 
signal and the slope of the spectral line wing, as given by 
Eq.\,(\ref{eqn:vi}), holds for both type of lines, so that the signals 
of emission and absorption lines add up rather than cancel. For 
simplification, we assume a typical Land\'e  factor of 
$g_{\rm eff} \approx 1.2$ for all lines.

The mean longitudinal magnetic field $\left<B_{\rm z}\right>$ is defined by the slope of the 
weighted linear regression line through the measured data points, where
the weight of each data point is given by the squared signal-to-noise ratio
of the Stokes $V$ spectrum. The formal $1\,\sigma$ error of 
$\left<B_{\rm z}\right>$ is obtained from the standard relations for weighted 
linear regression. This error is inversely proportional to the rms  
signal-to-noise ratio of Stokes $V$. Finally, the factor
$\sqrt{\chi^2_{\rm min}/\nu}$ is applied to the error determined from the 
linear regression, if larger than 1. Furthermore, we have carried out Monte Carlo bootstrapping tests. 
These are most often applied with the purpose of deriving robust estimates of standard errors. 
In these tests, we generate 250\,000 samples that have the same size as the original data set and analyse 
the distribution of the  $\left<B_{\rm z}\right>$ determined from all these newly generated data sets. 
The measurement uncertainties obtained before and after Monte Carlo bootstrapping tests were found to be 
in close agreement indicating the absence of reduction flaws. 

Additionally, to check the stability 
of the spectral lines along the full sequence of sub-exposures, we have compared 
the profiles of several spectral lines recorded in the parallel beam with the retarder waveplate at $+45^{\circ}$. 
The same was done for spectral lines recorded in the perpendicular beam. 
The line profiles looked pretty much identical within the noise.

\begin{table*}
\caption{
Logbook of the FORS\,2 polarimetric observations of the Of?p star CPD\,$-28^{\circ}$\,2561, including 
the modified Julian date of mid-exposure followed by the rotation phase, the open shutter time,
achieved signal-to-noise ratio, and the measurements of the mean longitudinal magnetic field using the 
Monte Carlo bootstrapping test, for all lines and excluding the lines in emission.
All quoted errors are 1$\sigma$ uncertainties.
Entries related to previous measurements are indicated by an 
asterisk in the first column of the Table. 
}
\label{tab:log_meas}
\centering
\begin{tabular}{lrrrr@{$\pm$}rr@{$\pm$}r}
\hline
\hline
\multicolumn{1}{c}{MJD} &
\multicolumn{1}{c}{Phase} &
\multicolumn{1}{c}{Shutter time} &
\multicolumn{1}{c}{S/N$_{5300}$} &
\multicolumn{2}{c}{$\left< B_{\rm z}\right>_{\rm all}$} &
\multicolumn{2}{c}{$\left< B_{\rm z}\right>_{\rm no\,emiss}$} \\
 &
 &
 \multicolumn{1}{c}{[s]} &
 &
 \multicolumn{2}{c}{[G]} &
 \multicolumn{2}{c}{[G]} \\
\hline
$^{*}$55338.969 & 0.453 & 3120 & 2262 & $-$388 & 102 & $-$340 & 122 \\ 
$^{*}$55685.982 & 0.181 & 5000 & 2481 &     47 &  82 &   $-$3 & 105 \\
$^{*}$55686.984 & 0.194 & 4800 & 2516 &  $-$47 &  78 &    -34 & 98 \\
$^{*}$55687.980 & 0.208 & 3800 & 2490 &    203 &  69 &    211 & 60 \\
      56631.229 & 0.057 & 460 &  214 &    525 & 564 &    520 & 566 \\
      56632.228 & 0.070 & 1840 & 1051 &    405 & 162 &    436 & 190 \\
      56640.258 & 0.180 & 2300 & 1136 &      0 & 145 &   $-$1 & 142 \\
      56641.292 & 0.194 & 2760 & 2087 &     37 &  69 &     53 & 70 \\
      56642.317 & 0.208 & 920 &  393 &     15 & 188 &    157 & 211 \\
      56643.181 & 0.220 & 2300 & 1418 &      6 & 111 &    152 & 94 \\
      56643.313 & 0.221 & 1800 & 1537 &     39 &  97 &    135 & 81 \\
      56644.192 & 0.231 & 2300 & 1201 &  $-$16 & 113 &      1 & 114 \\
      56646.114 & 0.260 & 2300 & 1272 &  $-$55 & 103 &     35 & 106 \\
      56647.128 & 0.273 & 2300 & 1322 & $-$134 &  78 &      0 & 104 \\
      56648.153 & 0.287 & 2300 & 1523 &     33 &  62 &  $-$35 & 61 \\
      56648.319 & 0.290 & 2300 & 1975 &  $-$32 &  71 &     17 & 65 \\
      56649.141 & 0.301 & 2300 & 1370 & $-$322 & 118 & $-$330 & 106 \\
      56649.316 & 0.303 & 2300 & 1810 & $-$247 &  87 & $-$205 & 73 \\
      56650.225 & 0.316 & 2720 & 1602 &  $-$84 &  94 &  $-$71 & 62 \\
      56656.272 & 0.398 & 2300 & 1777 & $-$345 &  90 & $-$348 & 96 \\
\hline
\end{tabular}
\end{table*}

\begin{figure}
\centering
\includegraphics[width=0.45\textwidth]{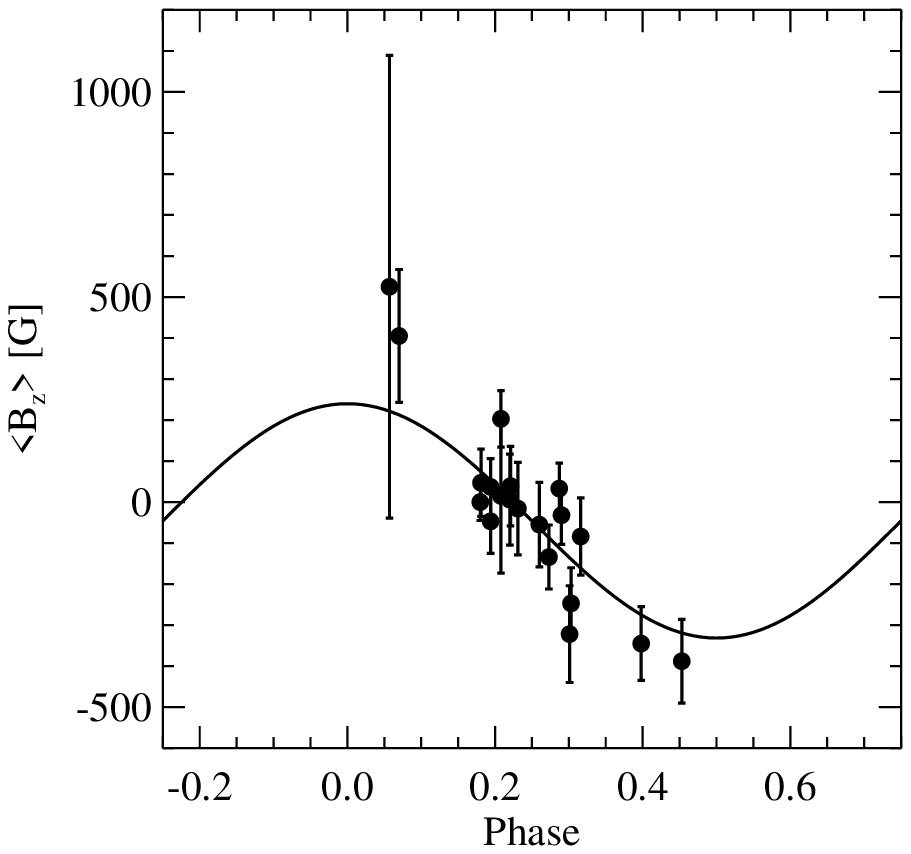}
\includegraphics[width=0.45\textwidth]{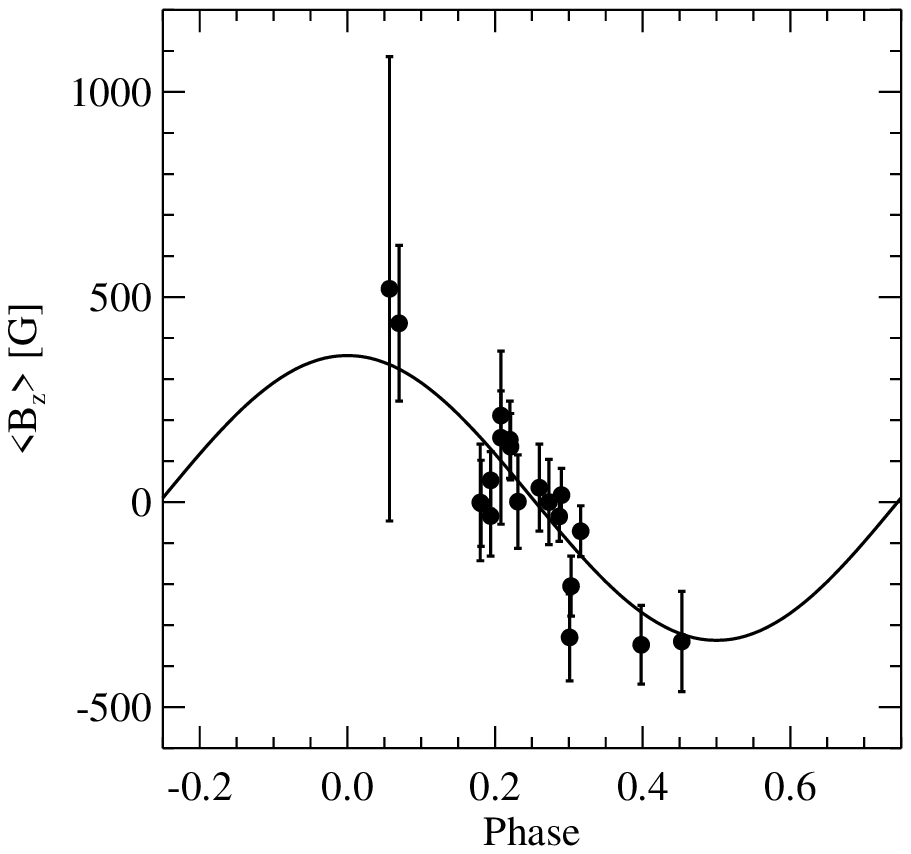}
\caption{
Longitudinal magnetic field variation of the Of?p star CPD\,$-28^{\circ}$\,2561 phased with the 73.41\,d 
rotation period reported by \citet{Naze2014}.
{\sl Top:}
Measurements of the magnetic field using absorption and emission lines.
The solid line represents a first fit to the data with a mean value of the magnetic field of
$\overline{\left<B_{\rm z}\right>} = -45.9\pm23$\,G and an amplitude of
$A_{\left<B_{\rm z}\right>} = 285.6\pm56$\,G.
{\sl Bottom:}
Measurements of the magnetic field using only the absorption lines.
The solid line represents a first fit to the data with a mean value of the magnetic field of
$\overline{\left<B_{\rm z}\right>} = 10.1\pm23$\,G and an amplitude of
$A_{\left<B_{\rm z}\right>} = 347.3\pm62$\,G.
For both fits, we assumed a rotation period of 73.41\,d and a zero phase at HJD2454645.49.
The accuracy of the fit is mainly limited by the insufficient phase coverage, which is not
even covering half the period.
}
\label{fig:rot}
\end{figure}

The results of our magnetic field measurements, both for absorption and emission lines combined
and absorption lines alone, are presented in 
Table~\ref{tab:log_meas}, where we also collect information about the modified Julian date,
the rotation phase, the open shutter time, and the achieved signal-to-noise ratio.
We note that the previous four measurements listed in Table~\ref{tab:log_meas} show now slightly
different values compared with the published ones \citep{Hubrig2011a,Hubrig2013}
since they have been re-analysed 
using the improved FORS\,2 data reduction described above. The longitudinal magnetic field variation of 
CPD\,$-28^{\circ}$\,2561 over the rotation period of 73.41\,d is presented in Fig.~\ref{fig:rot},
both for absorption and emission lines combined and absorption lines alone.

In spite of the fact that only a few measurements 
show 3$\sigma$ detections, 
the magnetic field variability with the rotation period of 73.41\,d is clearly detected in our data. 
The obtained results
confirm the excellent potential of FORS\,2 for measuring magnetic fields using low-resolution spectropolarimetry.
This potential was already demonstrated in the studies of the other two magnetic massive stars, $\theta^1$\,Ori\,C, 
which was the first O-type star with a detected weak magnetic field varying with a rotation period of 15.4\,days, and
the Of?p star HD\,148937 with a rotation period of 7.03\,d \citep{Hubrig2008,Hubrig2013}.

The smooth variation of $\left<B_{\rm z}\right>$ with phase in the FORS\,2 spectra of 
CPD\,$-28^{\circ}$\,2561 indicates that the rotation period is indeed very long. 
A frequency analysis applied to our magnetic data (see e.g., \citealt{Hubrig2011c}) revealed a prominent 
frequency corresponding to a period of 85.6\,d. A smaller peak at the frequency corresponding to
a period of 73.41\,d coincided with the
window function and therefore is indistinguishable from an artifact.

The observed  single-wave variation in the longitudinal magnetic field during the stellar rotation cycle indicates
a dominant dipolar contribution to the magnetic field topology. Assuming that the star is an oblique dipole rotator, 
the magnetic dipole axis tilt $\beta$ is constrained by

\begin{equation}
\frac{\left<B_{\rm z}\right>_{\rm max}}{\left<B_{\rm z}\right>_{\rm min}} =\frac{\cos(i + \beta)}{\cos(i - \beta)},
\end{equation}

\noindent
where the inclination
angle $i$ can be derived from considerations of the stellar fundamental parameters.
Using for the stellar radius $R=14\,R_{\odot}$ \citep{Petit2013}
combined with the period of 73.41\,d we obtain $v_{\rm eq}=9.65$\,km\,s$^{-1}$.
On the other hand, it is clear that it is not possible to use the low-resolution FORS\,2 spectra to estimate 
the $v\,\sin\,i$ value so that the inclination angle $i$ is not known and only a low limit of dipole strength 
$B_d$ can be estimated. We note, however, that the spectral behaviour of CPD\,$-28^{\circ}$\,2561 is highly variable
over the rotation period (see Sect.~\ref{sect:var}), indicating that the angle $i$ cannot be close to zero. 
Taking into account the parameters of the sinusoids fitted to our measurements and
assuming $\left<B_{\rm z}\right>_{\rm max}\approx350$\,G and a limb-darkening coefficient $u=0.4$,
we conclude that the polar strength of the surface dipole $B_d$ is larger than 1.15\,kG.

\section{Spectral variability and wind parameters of CPD\,$-28^{\circ}$\,2561}
\label{sect:var}

Among the massive star samples studied by \citet{Hubrig2011a,Hubrig2013}, both Of?p stars 
CPD\,$-28^{\circ}$\,2561 and HD\,148937 exhibited a most conspicuous spectral variability detectable in low-resolution
FORS\,2 spectra. 
With a total of 20 spectropolarimetric observations of CPD\,$-28^{\circ}$\,2561 on our disposal it became 
now possible to investigate 
the spectral variability of this star in more detail over approximately half of the rotation cycle. 

In the following, we discuss the behaviour of several diagnostic lines in the spectra of this 
star in comparison with HD\,148937 for which we carried out seven observations with FORS\,2 during the last
years. Both stars have approximately the same spectral type around O6, and exhibit a similar 
spectral appearance. The first detection of a mean longitudinal 
magnetic field in the Of?p star HD\,148937, $\langle B_{\rm z}\rangle=-254\pm81$\,G,
using FORS\,1 at the VLT was reported by \citet{Hubrig2008}.

For a better presentation of the behaviour of different lines in the spectra of CPD\,$-28^{\circ}$\,2561 at different 
rotation phases,
we grouped the observations in four phase intervals, namely the observations obtained in the phase 
intervals from 0 to 0.12, from 0.12 to 0.24, from 0.24 to 0.36, and from 0.36 to 0.46.
As in the work of \citet{Hubrig2013}, to calculate the rotation phase of 
the observed spectra of HD\,148937, we used the rotation period of 7.03\,d determined by \citet{Naze2008}, who
also mentioned a lower-amplitude variability of this star.

\begin{figure}
\centering
\includegraphics[width=0.65\textwidth]{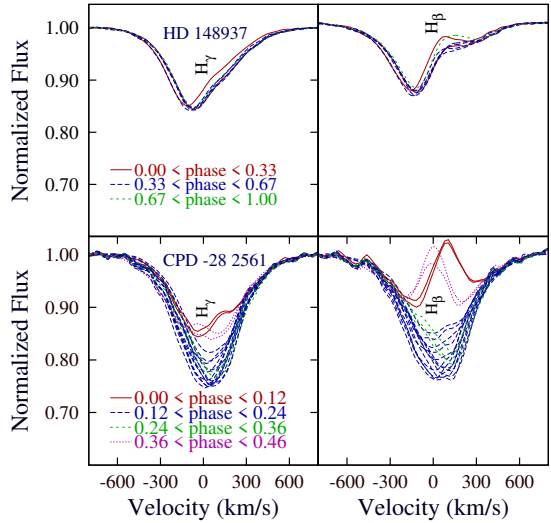}
\caption{
The behaviour of the hydrogen lines H$\beta$ and H$\gamma$ phased  with the rotation period
of 73.41\,d (lower panel) compared to the behaviour of the hydrogen lines in HD\,148937 (upper panel).
}
\label{fig:hydr}
\end{figure}

The comparison of the spectral behaviour of CPD\,$-28^{\circ}$\,2561 with that of HD\,148937
reveals a remarkable difference in the degree of variability between both stars.
In Fig.~\ref{fig:hydr}, we compare the behaviour of the lower number hydrogen lines H$\gamma$ and H$\beta$ 
phased with the rotation period of 73.41\,d with that of the hydrogen lines in the spectra of HD\,148937.
Opposite to the behaviour of the hydrogen lines in HD\,148937, the H$\gamma$ and H$\beta$ lines in the spectrum of 
CPD\,$-28^{\circ}$\,2561 undergo significant profile changes over half of the rotation cycle with the strongest 
contribution of the emission component at positive and negative extrema of the magnetic field.

\begin{figure}
\centering
\includegraphics[width=0.45\textwidth]{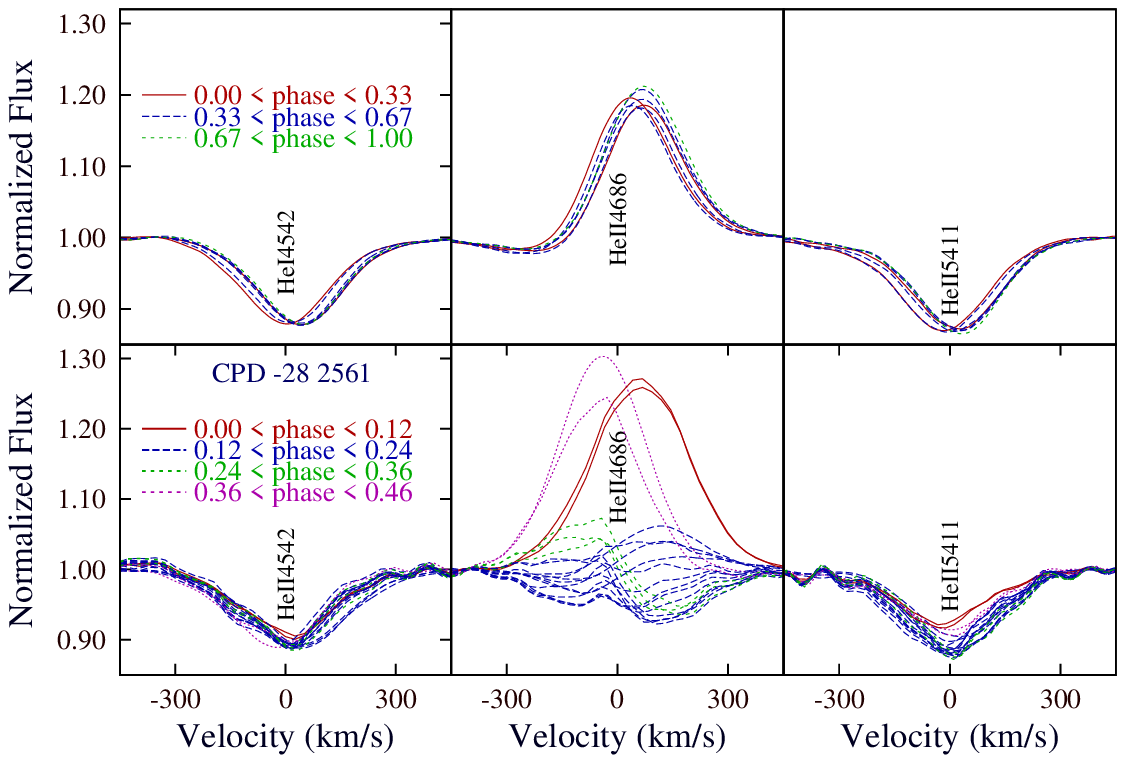}
\caption{The behaviour of  the \ion{He}{ii} $\lambda$4542, \ion{He}{ii} $\lambda$4686 and  
\ion{He}{ii} $\lambda$5411 lines phased  
with the rotation period of 73.41\,d (lower panel) compared to the behaviour of the  
\ion{He}{ii} lines in HD\,148937 (upper panel).
}
\label{fig:he2}
\end{figure}

A very similar behaviour to that of hydrogen lines is observed for the \ion{He}{ii} $\lambda$4686 line,
which appears in full emission in the phases of the magnetic field extrema.
Other \ion{He}{ii} lines show no distinct emission component, but it cannot be excluded that they are partly 
filled in by stellar-wind emission
in the phases coinciding with the strongest magnetic field strength.
The behaviour of \ion{He}{ii} $\lambda$4542, \ion{He}{ii} $\lambda$4686 and  \ion{He}{ii} $\lambda$5411 
lines phased with the 
rotation period of 73.41\,d, in comparison to that of the  \ion{He}{ii} lines in in the spectra of 
HD\,148937, is presented in Fig.~\ref{fig:he2}.

\begin{figure}
\centering
\includegraphics[width=0.45\textwidth]{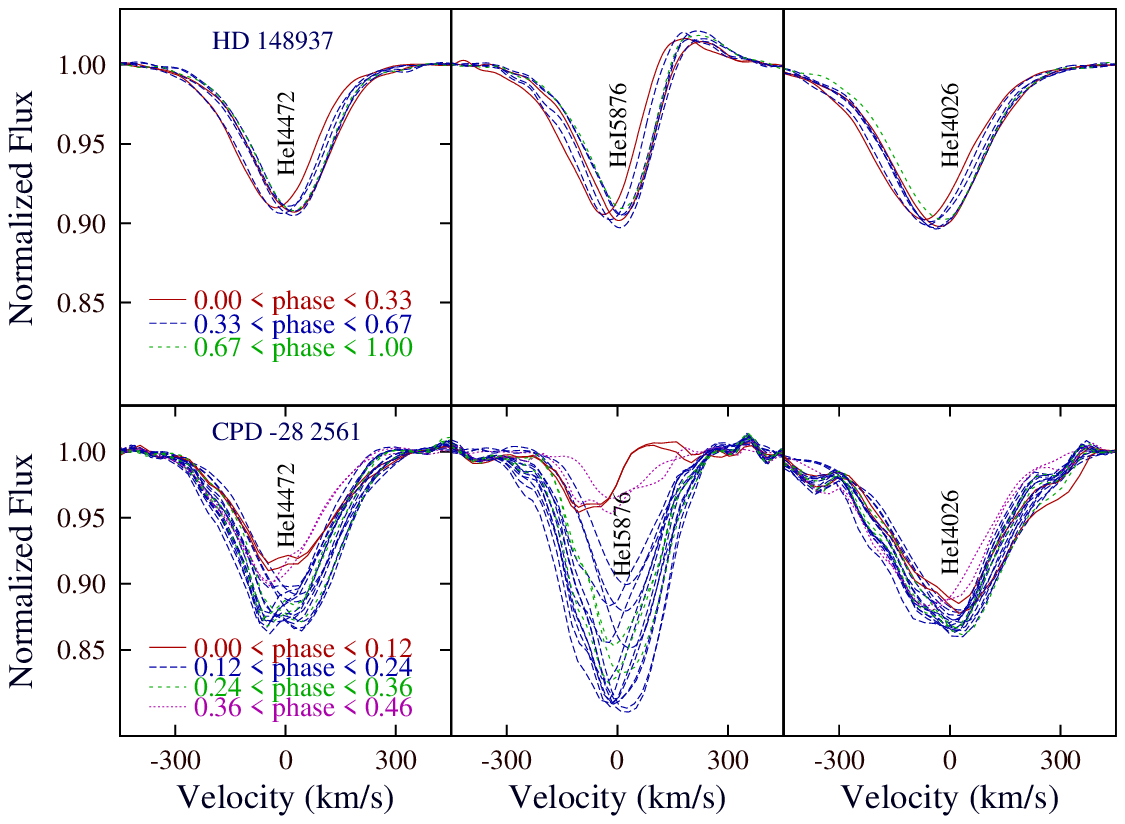}
\caption{The behaviour of the \ion{He}{i} $\lambda$4472, \ion{He}{i} $\lambda$5876, and \ion{He}{i} $\lambda$4026,
phased with the rotation period of 73.41\,d (lower panel) compared to the behaviour of the  \ion{He}{i} lines in HD\,148937 (upper panel). 
}
\label{fig:he1}
\end{figure}

Among the \ion{He}{i} lines, the strongest variation with rotation phase is detected in \ion{He}{i} $\lambda$5876,
followed by the \ion{He}{i} $\lambda$4472 line. Interestingly, as is shown in Fig.~\ref{fig:he1}, the behaviour 
of the emission component in the core of the \ion{He}{i} line
is reminiscent of that visible in the core of the H$\gamma$ line presented in Fig.~\ref{fig:hydr}. 

\begin{figure}
\centering
\includegraphics[width=0.40\textwidth]{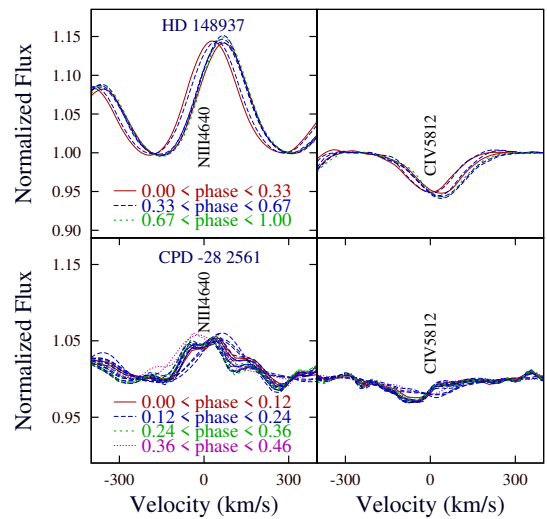}
\caption{The behaviour of the metal lines \ion{N}{iii} $\lambda$4640 and  \ion{C}{iv} $\lambda$5812 phased  
with the rotation period of 73.41\,d (lower panel) compared to the behaviour of the same metal lines in HD\,148937 
(upper panel). 
}
\label{fig:metal}
\end{figure}

Also the metal lines exhibit distinct variability over the rotation period. The absorption component in the metal 
lines seems to decrease at the phases of the magnetic positive and negative extrema. 
The variability of the metal lines \ion{N}{iii} $\lambda$4640 and  \ion{C}{iv} $\lambda$5812 phased  
with the rotation period of 73.41\,d is presented in Fig.~\ref{fig:metal}. 
This trend in the behaviour of metal 
lines is similar to the trend found for hydrogen and helium lines. 

In summary, CPD\,$-28^{\circ}$\,2561 shows a much stronger spectral variability compared to HD\,148937.
According to the Magnetic Wind Confinement (MWC) model by \citet{Babel1997}, dipolar magnetic fields 
in magnetic massive stars are able
to guide stellar winds from the two opposite hemispheres towards the magnetic equator where the flows collide
and the colliding flows shock to create a high-temperature, X-ray-emitting plasma. The plasma cools to 
form a dense disk that gives rise to optical emission.
Since the reason for the variability in magnetic massive stars is usually referred to the 
varying projection angle of a plasma disk confined to the magnetic equatorial plane, it is possible that 
HD\,148937 is viewed at relatively low inclination compared to the axial inclination of CPD\,$-28^{\circ}$\,2561.
Alternatively, the inclination angle could be large while the obliquity between the magnetic field
axis and the rotation angle is small.

\begin{figure}
\centering
\includegraphics[width=0.42\textwidth]{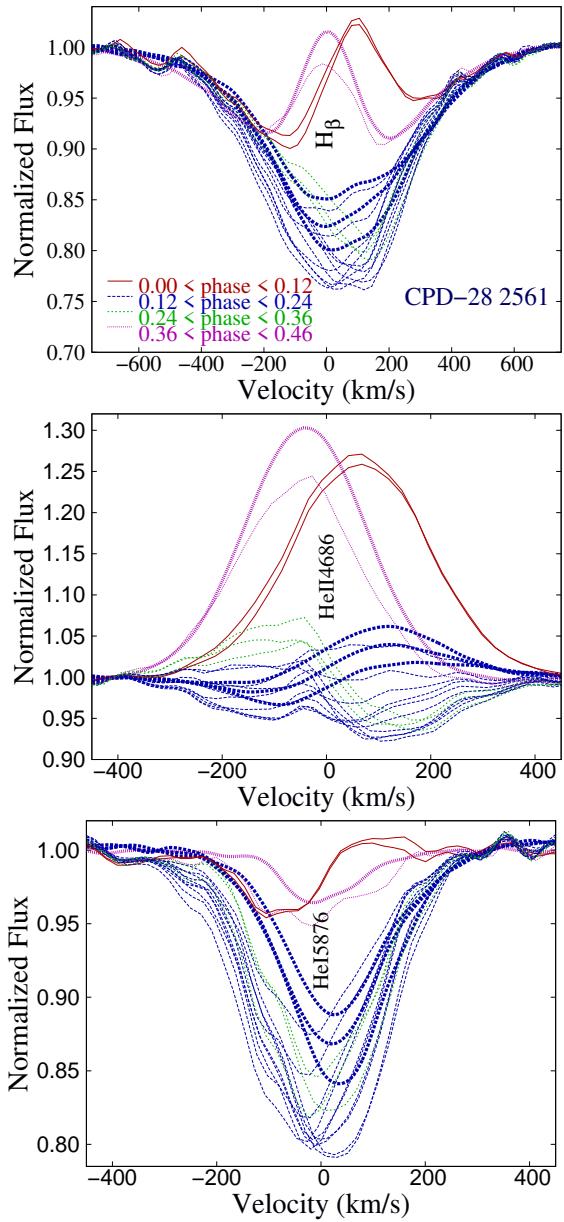}
\caption{
Hydrogen and helium line profiles observed in the years 2010--2013.
The profiles observed in the years 2010 and 2011 are highlighted by thicker lines.
}
\label{fig:civ}
\end{figure}

During the study of the behaviour of the line profile shapes belonging to different elements,
we detected notable differences in the 
line profiles observed in identical rotation phases in the years 2010 and 2011 compared to 
those observed in 2013 December.
As an example, we present in Fig.~\ref{fig:civ} the profiles of the lines H$\beta$, 
\ion{He}{II} $\lambda$4686, and \ion{He}{I} $\lambda$5876, 
observed from 2010 to 2013, with observations in the years 2010 and 2011 highlighted by thicker lines.
These differences can probably be attributed either to the inaccuracy of the 
determination of 
the rotation period or to the intrinsic variability of this star, probably due to changes in the 
amount or distribution of emitting material with time in the magnetosphere of CPD\,$-28^{\circ}$\,2561.
Such a phenomenon has already been reported for other well-studied Of?p stars (\citealt{Howarth2007}).

\begin{figure}
\centering
\includegraphics[width=0.39\textwidth]{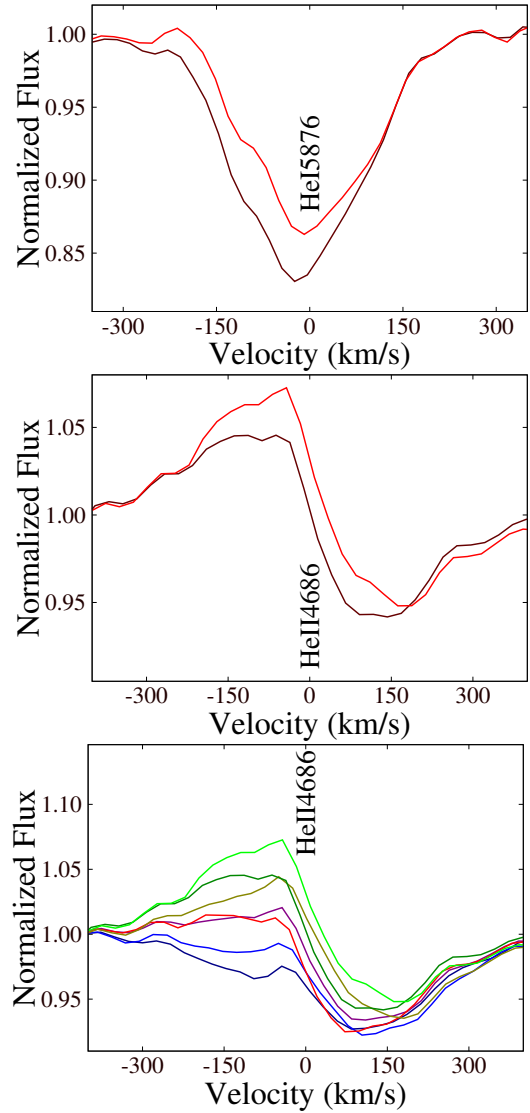}
\caption{
{\it Upper and middle panel}: Short-term spectral variability of the \ion{He}{i} $\lambda$5876 and \ion{He}{ii} $\lambda$4686 
lines observed on the night of 2013 December 23 separated by $\sim$4 hours.
{\it Lower panel}: Day-to-day variations observed in the line profile of \ion{He}{ii} $\lambda$4686
during the nights from 2013 December 20 to 24.
 }
\label{fig:day}
\end{figure}

To assess the short-term spectral variability of lines belonging to different elements,
spectropolarimetric observations have been carried out at time intervals of just a 
few hours on several nights. The short-term spectral variability is easily detectable
in several diagnostic lines presented in Fig.~\ref{fig:day}, indicating that these changes in the line 
profiles of CPD\,$-28^{\circ}$\,2561 cannot be explained exclusively by rotational or magnetic modulation.
This kind of variability is probably rather related to intrinsic variability of CPD\,$-28^{\circ}$\,2561.

\begin{figure*}
\centering
\includegraphics[width=0.30\textwidth]{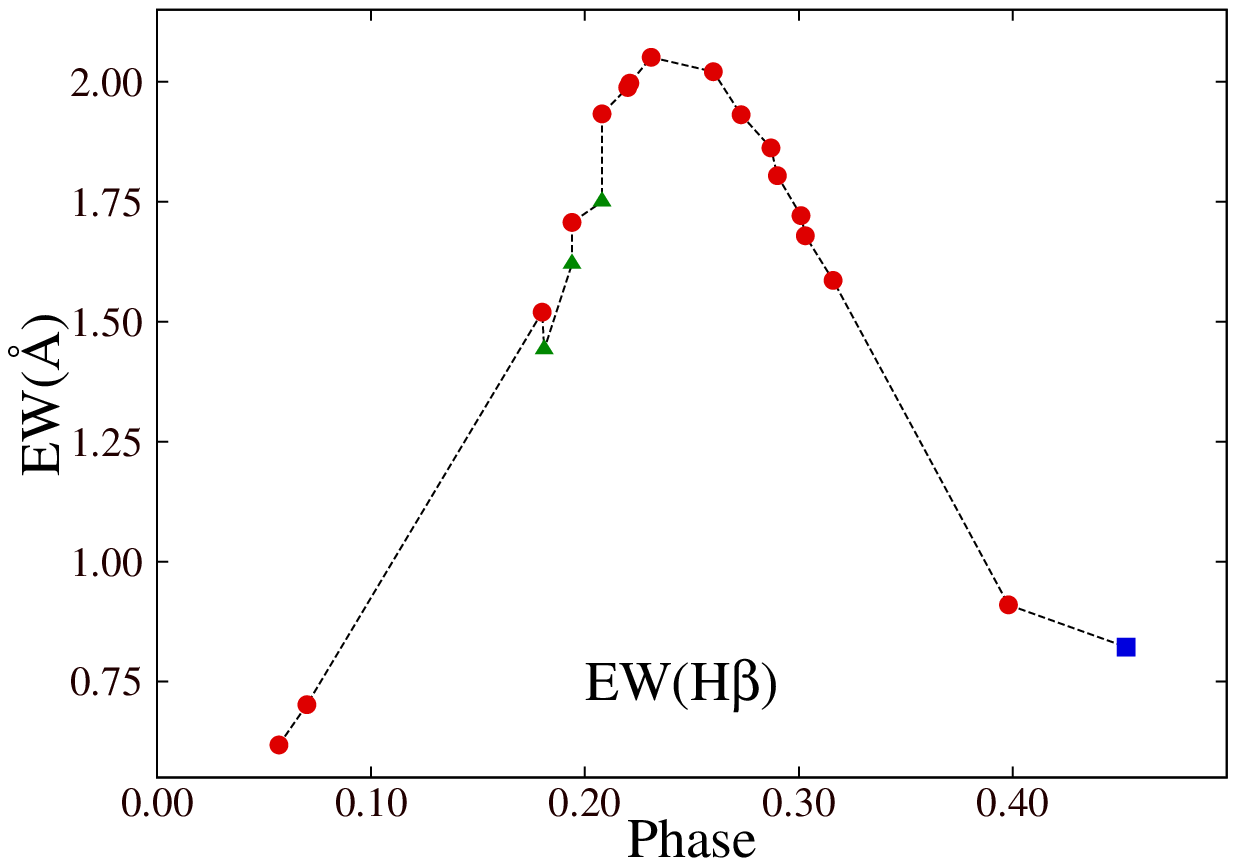}
\includegraphics[width=0.30\textwidth]{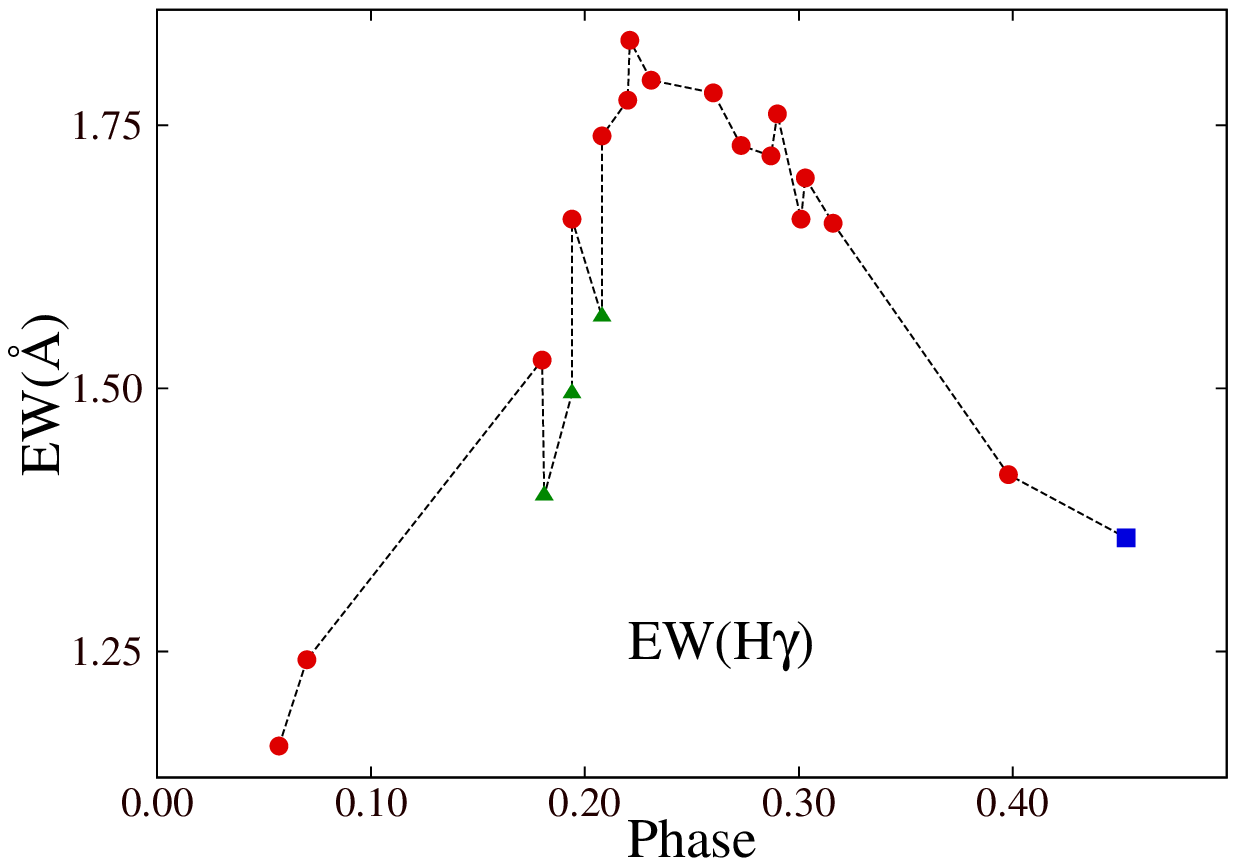}
\includegraphics[width=0.30\textwidth]{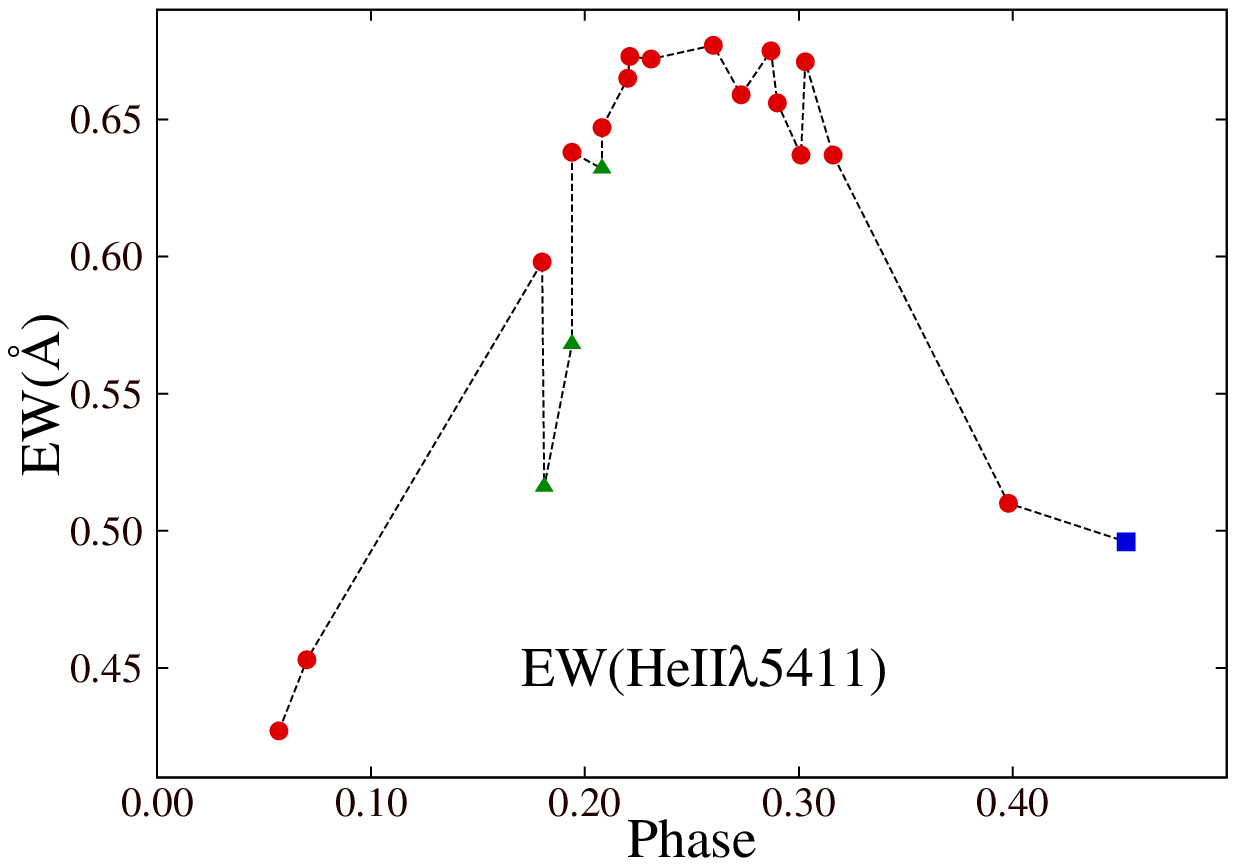}
\includegraphics[width=0.30\textwidth]{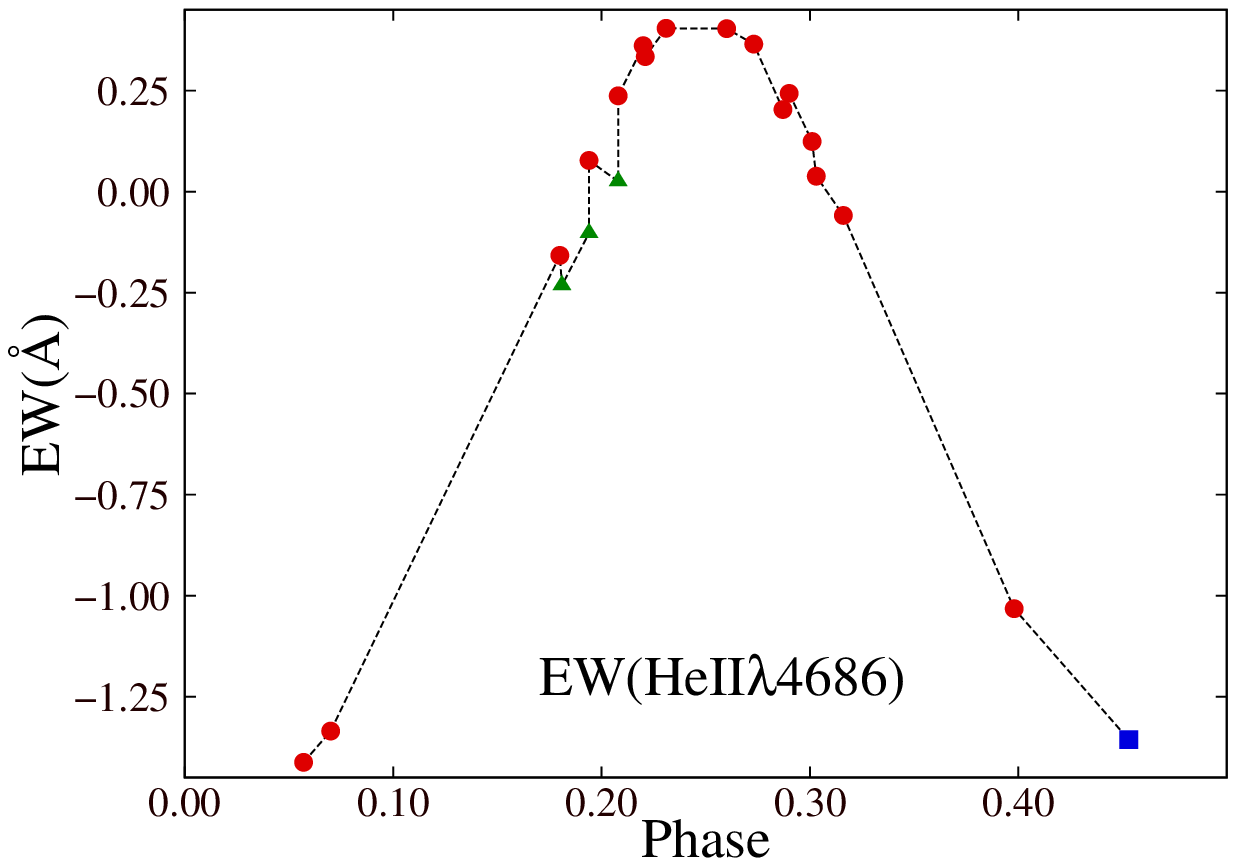}
\includegraphics[width=0.30\textwidth]{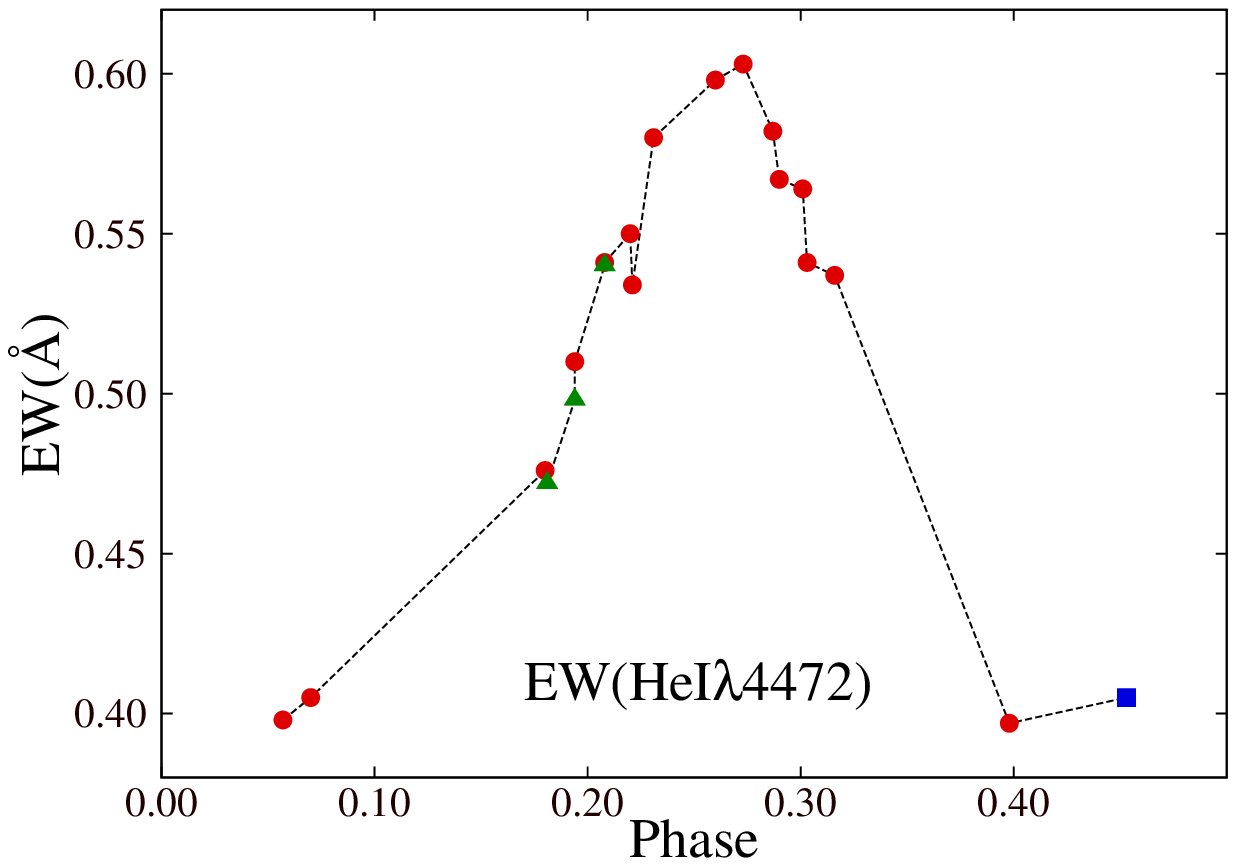}
\includegraphics[width=0.30\textwidth]{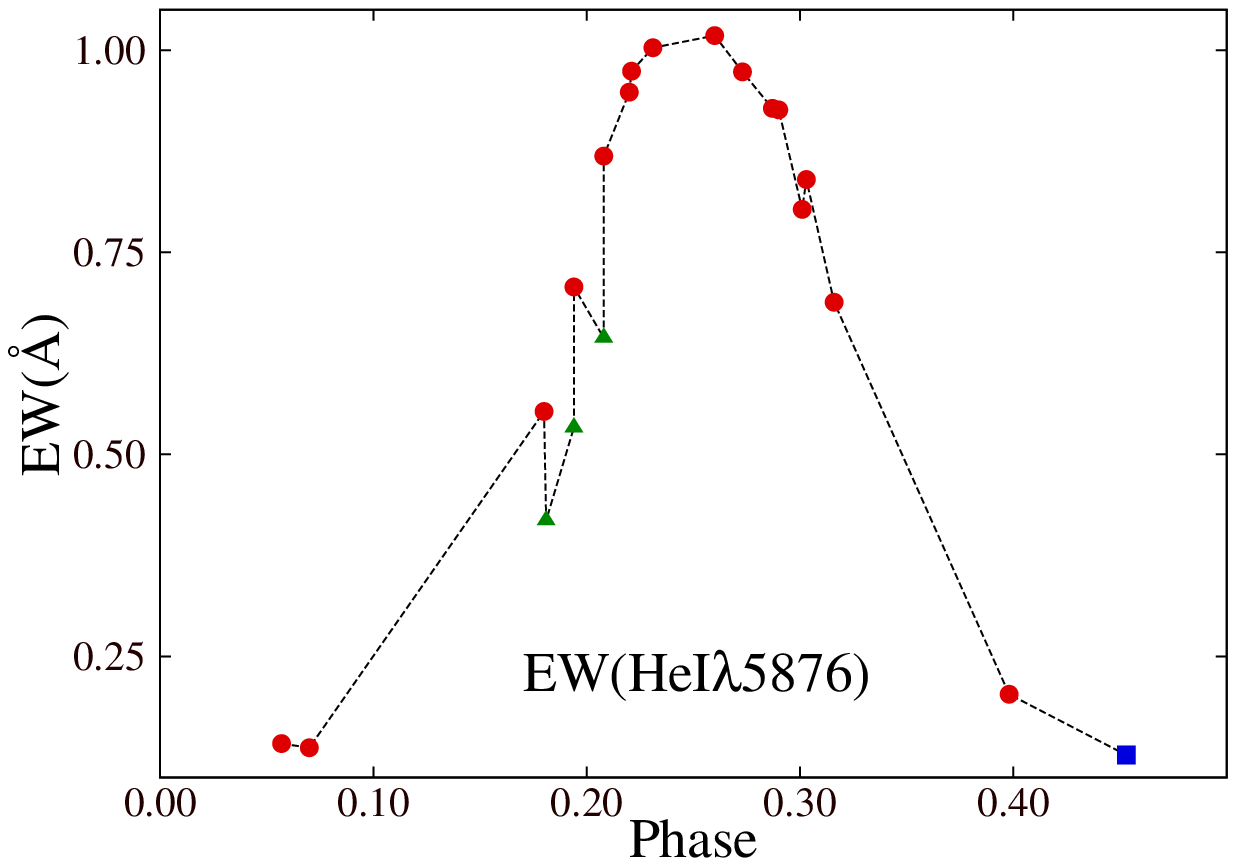}
\includegraphics[width=0.30\textwidth]{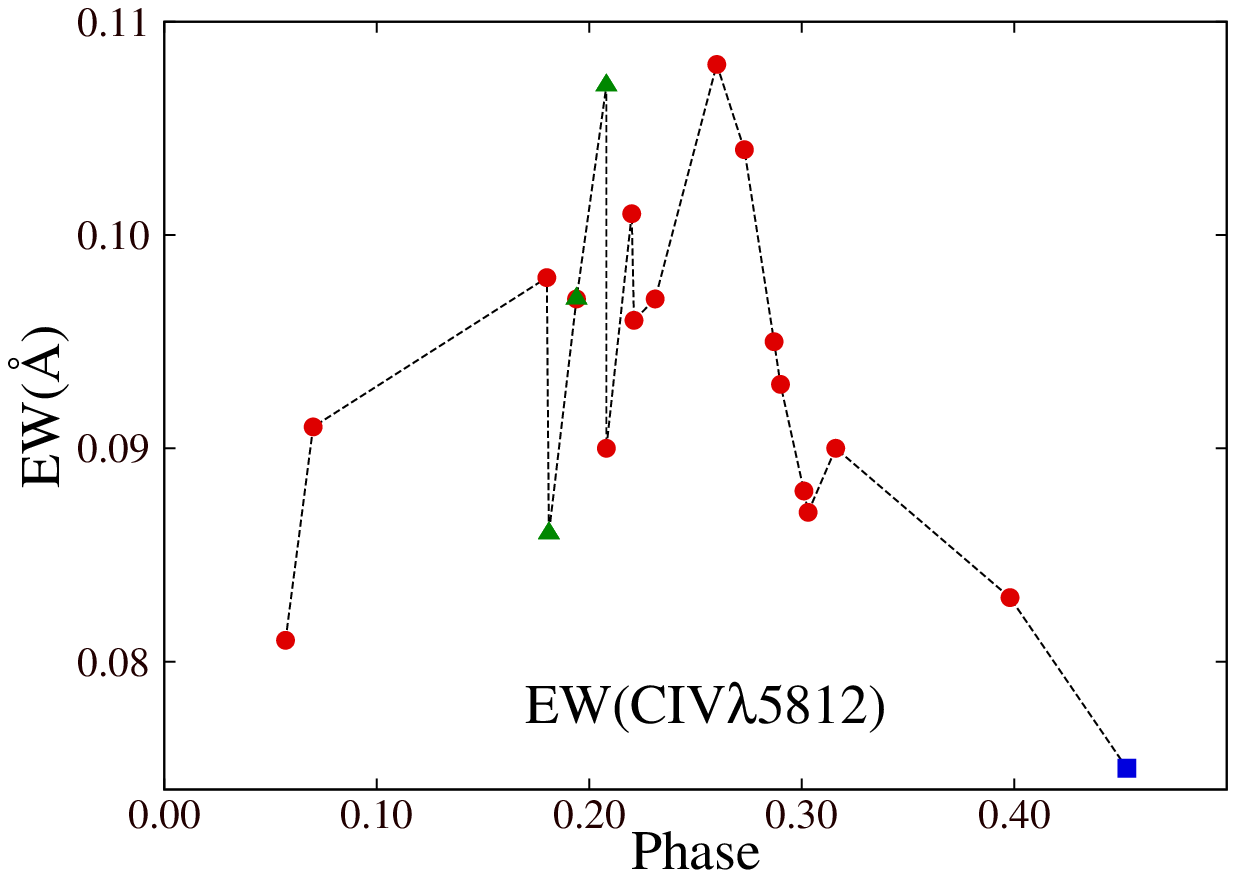}
\includegraphics[width=0.30\textwidth]{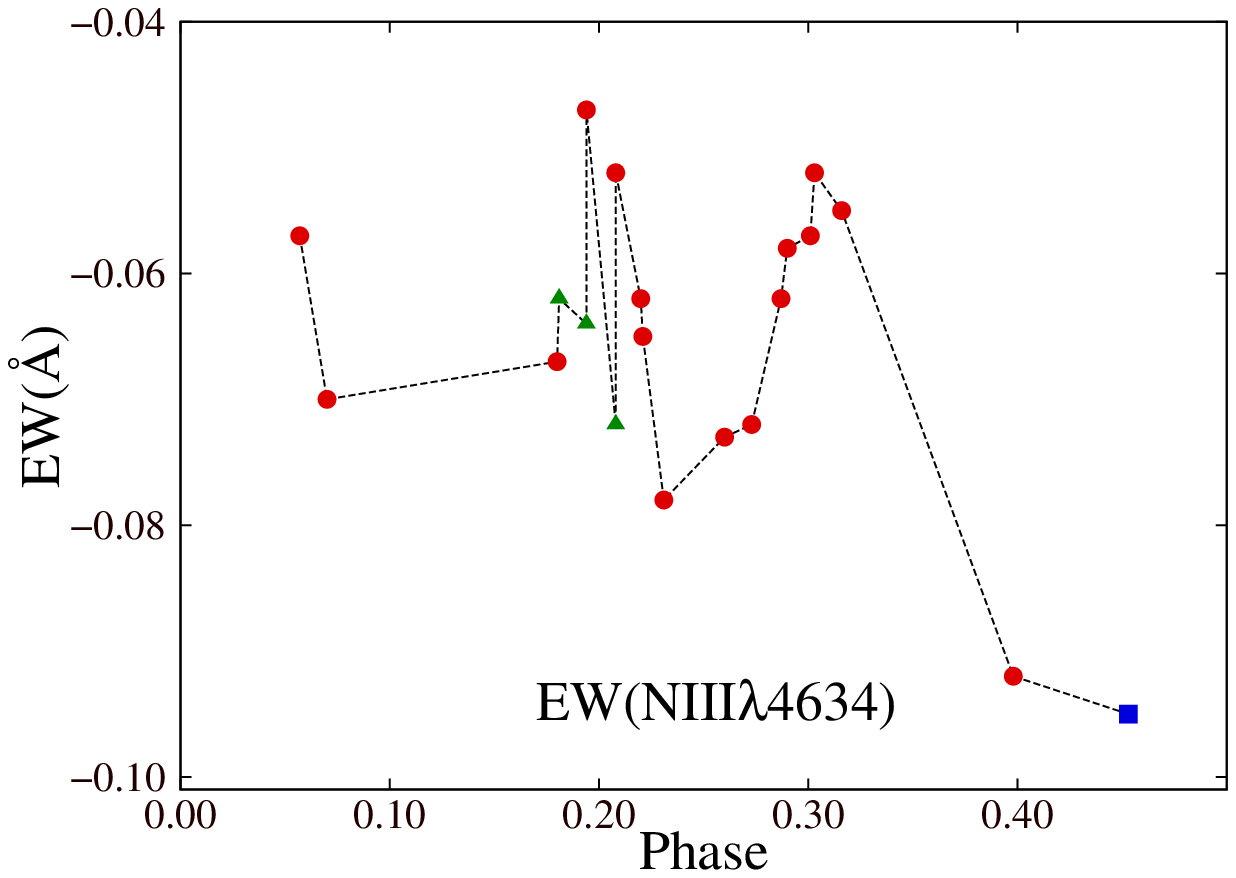}
\includegraphics[width=0.30\textwidth]{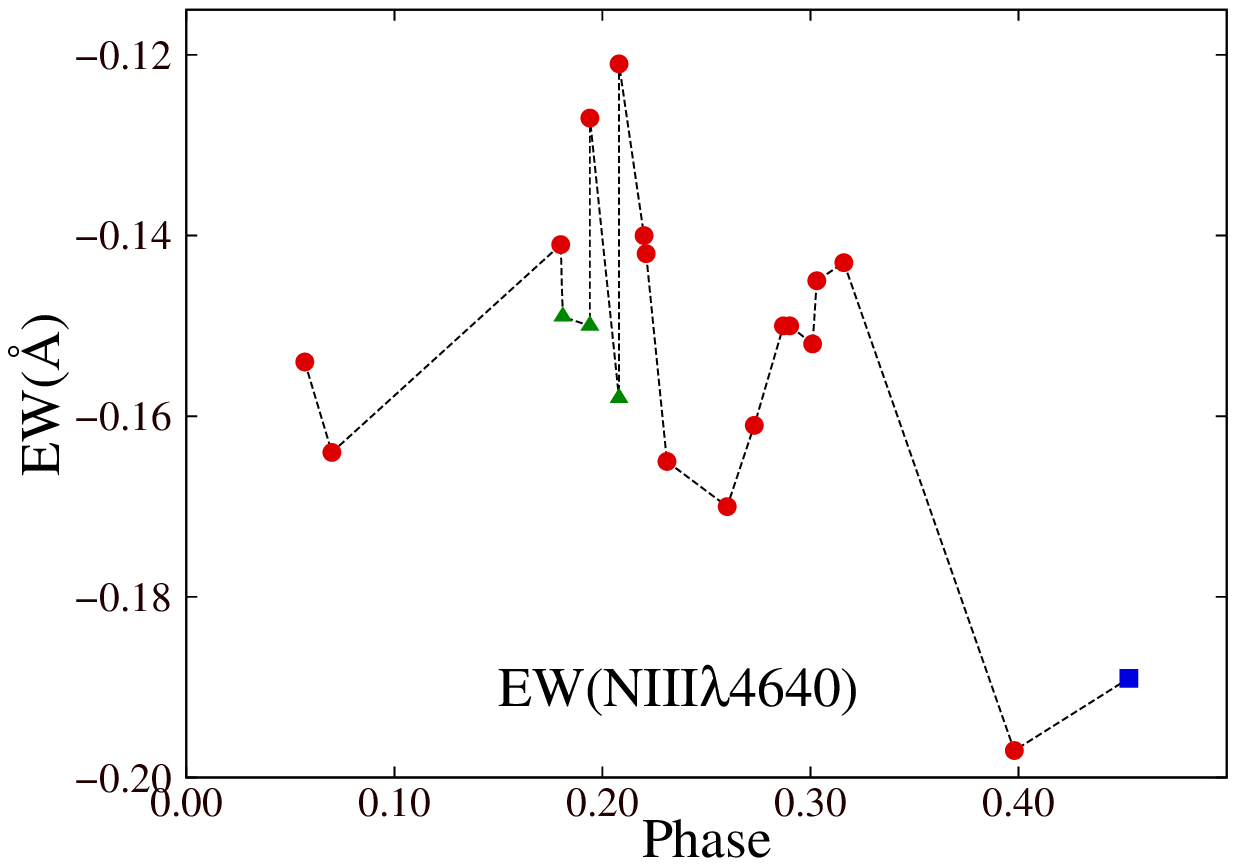}
\caption{
The phased equivalent width measurements of the H$\beta$ and H$\gamma$ lines, the \ion{He}{ii} lines $\lambda$5411 and 
$\lambda$4686, the \ion{He}{i} lines $\lambda$4472 and $\lambda$5876,
and the metal lines \ion{C}{iv} $\lambda$5812, \ion{N}{iii} $\lambda$4634, and \ion{N}{iii} $\lambda$4640.
The error bars are very small, of the order of the size of the filled circles. 
They were estimated from rms scatter in the continuum region.
Blue squares correspond to measurements from 2010, green triangles to measurements from 2011,
and red circles to measurements from 2013.
}
\label{fig:ew}
\end{figure*}

Finally, we present in Fig.~\ref{fig:ew} equivalent width variations of several spectral lines phased with the 
rotation period of 73.41\,d. The variation of all studied lines is approximately in phase with the 
variations of the hydrogen lines. 
The equivalent widths were obtained by numerically integrating over the 
normalized line profiles. 
The uncertainties were estimated from the rms scatter in the continuum regions around the line profiles. 
The large scatter of the measurements around the rotation phase 0.2 is caused by differences in the behaviour of the line
profiles recorded in 2010--2011 and those recorded in 2013. The behaviour of the equivalent width phased curve for
\ion{N}{iii} $\lambda$4640 indicates the possible presence of a second minimum around the phase 0.25.
More observations are needed to investigate the behaviour of nitrogen in more detail.

\begin{figure}
\centering
\includegraphics[width=0.45\textwidth]{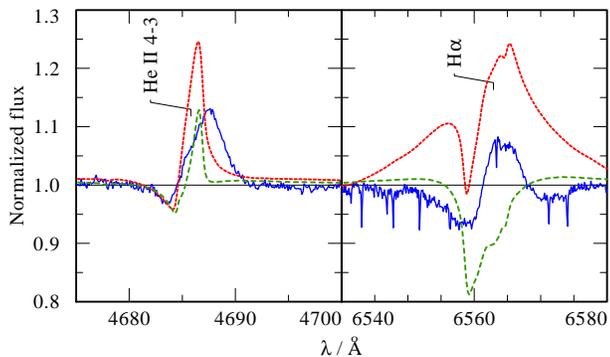}
\caption{
Details of the optical spectrum of CPD\,$-28^\circ$\,2561: Observation (blue solid line) vs.\ a
model with $\log \dot{M} = -6$
(red dotted) and a model with $\log \dot{M} = -5.8$ (green dashed), both
with $v_\infty=1000\,$km\,s$^{-1}$ and stellar parameters as described in the text.
 }
\label{fig:model}
\end{figure}

As no UV spectrum of CPD\,$-28^\circ$\,2561 is currently available, 
to infer the mass-loss rate $\dot{M}$ and the terminal velocity $v_\infty$ of the wind of CPD\,$-28^\circ$\,2561,
we tried to estimate these parameters
from a high-resolution optical spectrum only.
We downloaded from the ESO archive the publically available 
high-resolution spectropolarimetric observation of  CPD\,$-28^\circ$\,2561 obtained with the 
HARPS polarimeter (HARPSpol; \citealt{Snik2008}) attached to ESO's 3.6 m telescope (La Silla, Chile) on 2011 December
10, which corresponds to the rotation phase 0.17.
The obtained spectropolarimetric observation has a very low signal-to-noise ratio around 100 in the Stokes~$I$ spectrum 
and a resolving power of about $R = 115\,000$, covering the spectral range 3780--9107\AA{}, 
with a small gap between 5259 and 5337\AA{}.  We note that such a low signal-to-noise ratio of this HARPS spectrum
does not allow us to use the spectrum for the measurement of the magnetic field.
The reduction and calibration of the archive spectrum were
 performed using the HARPS data reduction software available at the ESO headquarter in Germany.

However, we already noted from Figs.~\ref{fig:hydr}--\ref{fig:day} the high spectral variability of the hydrogen 
and helium lines.
As an example, the \ion{He}{II} $\lambda$4686 line (Fig.~\ref{fig:civ}) changes from emission to absorption
and the center of the line profiles changes from blueshifted to redshifted.
It is therefore unlikely that the observed emission line profiles originate from a stationary symmetric wind.
A stationary asymmetric, e.g.\ polar, outflow is also not very likely, as the observed time 
scales of the line profile variability is much shorter than the rotation period assumed.
Moreover, we found that the stellar parameters of CPD\,$-28^\circ$\,2561, given in \citet{Petit2013}, are inconsistent.
For the given luminosity, $T_{\rm eff}$, and  $\log g$, the stellar mass should be 86\,M$_\odot$.
However, \citet{Petit2013} give 43\,M$_\odot$. 
Since we do not know which distance was used in \citet{Petit2013},
we adopt in our analysis the spectroscopic distance of 6.1\,kpc as determined by  
\citet{Patri2003}.

Therefore, instead of an accurate reproduction of the optical spectrum, to get an impression of the impact of a 
spherical symmetric wind with 
typical O star wind parameters on the emission lines, such as \ion{He}{II} $\lambda$4686 and H$\alpha$,
we calculated a set of {\sc PoWR} model atmospheres 
\citep{graefener2002,hamgrae2004}.
{\sc PoWR} solves the radiative transfer equation in a co-moving frame and performs full
non-LTE calculations of population numbers.
Accordingly, it is appropriate for expanding atmospheres.
We adopted solar abundances and $T_{\rm eff}$, $\log L$, 
and $\log g$ from \citet{Petit2013} and tried different combinations of $\dot{M}$ and $v_\infty$.
Our tentative analysis indicates that for 
the given stellar parameters a symmetric wind with a typical  $\dot{M}$ of about $10^{-6}\,{\rm M}_\odot$/yr
(cf.\ \citealt{vink2001}) would 
lead to emission lines (see for example  Fig.~\ref{fig:model} presenting \ion{He}{II} $\lambda$4686),
which are not seen for all phases of the rotation.
Further tests indicated that some of the emission features can be reproduced by a slow outflow with velocities of 
$v_\infty<500$\,km\,s$^{-1}$, 
which is much below the terminal velocities of other O stars (cf.\ \citealt{puls1996}).
Thus, it seems to be likely that the 
O star wind is strongly affected by the magnetic field and the observed variable emission features originate neither 
from a stationary symmetric O star wind, nor from a polar outflow modulated with the rotational period.
It can only be 
speculated if these emission features may stem from a highly variable
magnetosphere as described by \citet{udo2013}.  

The magnetic field dominates  
the stellar wind below the radial height $R_{\rm w}$, where the condition
$\frac{B^{2}}{8\pi} < \frac{\rho v^2}{2}$
is met, while above this height the solar wind is capable to stretch
and open the magnetic field lines \citep{weber1967,AltschulerNewkirk1969}.
According to our spectropolarimetric measurements,
the magnetic field in CPD\,$-28^{\circ}$\,2561 is strong
enough to affect the stellar wind up to $R_{\rm w} \sim 4\,R_\ast$ when we use 
the stellar wind parameters estimated from our modeling and $\rho=\dot{M}/4\pi r^2 v_\infty$. 
Since the hydrogen Balmer
lines are formed in the lower parts of the wind, the observed spectral
variability in those lines may be understood as being due to the wind
material affected by the magnetic field.
Recently, a first attempt to explain
the spectral variability in Of?p stars was made by \citet{sund2012}.
These simulations rely on the assumption of a local thermodynamic
equilibrium (LTE) in the Of?p-star wind that is not realistic, as
O-star winds are dominated by non-LTE processes.
Nevertheless, their effort highlights the potential importance of non-spherical geometries
in strongly magnetic O-stars for the detailed analysis of line-profile
variability.

\section{X-ray observations with the {\em Suzaku X-ray Observatory}}
\label{sect:xrays}

CPD\,$-28^{\circ}$\,2561 is an X-ray source detected by the {\em Rosat} X-ray
observatory in 1990. From the ROSAT All-Sky Survey Faint Source Catalogue
(RASS-FSC) its count rate is $0.03\pm 0.01$\,s$^{-1}$.
A radio source is
also located at the same position as the X-ray source \citep{fh2004}.
Already {\em Rosat} observations indicated that the ratio of X-ray
to bolometric luminosity ($\log{L_{\rm X}/L_{\rm bol}}\approx -6$)
makes CPD\,$-28^{\circ}$\,2561 X-ray brighter compared to an average O-type
star \citep{Chlebowski1989,Berg1997,Oskinova2005}.
The {\em Rosat} PSPC hardness ratio (${\rm HR1}\approx 0.04\pm 0.4$) provided
some evidence that X-ray emission from CPD\,$-28^{\circ}$\,2561 can be harder
than is typical for an O-type star.

To study its X-ray properties in more detail, we observed
CPD\,$-28^{\circ}$\,2561 with the {\em Suzaku X-ray Observatory}
\citep{Mitsuda2007} on 2014 April 28 for 27~ksec. 
Data were obtained with the XIS0, XIS1, and XIS3 detectors on board
{\em Suzaku}. The star is confidently detected in all detectors.
Standard data analysis using the most recent calibration
was performed.

\begin{figure}
\centering
\includegraphics[width=0.45\textwidth]{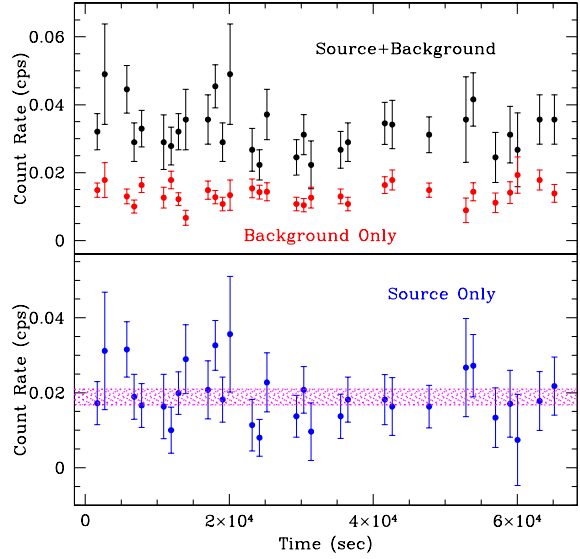}
\caption{
The X-ray light curve of CPD\,$-28^{\circ}$\,2561.
{\it Upper panel}:  The X-ray light-curve from the region with $2^{\prime}$
radius centered on the position of CPD\,$-28^{\circ}$\,2561 without subtracting the
background (upper black data points) and the X-ray light-curve from a
background region (lower red data points).
{\it Lower panel}: The
source count rate is presented with a binning of 1024\,s, after subtracting the
background
count rate.  The magenta band represents the $\pm 1 \sigma$
region about the average value for these data points.
}
\label{fig:Xlightcurve}
\end{figure}

To investigate the X-ray variability, we produced the X-ray light-curves using
different time binning.
Data from all three detectors in the energy range of 0.7 to
2.0\,keV have been combined and the counts within a circle of $2^{\prime}$
radius centered on the source position were collected. An annulus around the
source position with an outer radius of $4^{\prime}$ was used as a
background region. The X-ray background light-curve
was extracted from this region. Figure~\ref{fig:Xlightcurve} shows the resulting
X-ray light curve using a binning of 1024 seconds.
 
The resulting light-curve was tested for variability. We tested the hypothesis
of a constant source (see lower panel in Fig.~\ref{fig:Xlightcurve}) using
$\chi^2$ fitting. A value of $\chi^2 = 25.2$ with 30 degrees of freedom (``dof'')
indicates that the hypothesis of a constant source can {\em not} be rejected at
better than 90\% confidence.  We conclude that there is no evidence
for X-ray variability from this single observation with a total
duration of about 7.5 hours. The rotation period for CPD\,$-28^{\circ}$\,2561
is about 73.4 days.  Thus our pointing covers less than half a percent of
the star's rotation (and its associated magnetosphere).
We therefore cannot make any conclusion concerning
cyclic X-ray variations over the star's rotation period.
High amplitude and rapid changes relative to the rotation period are not detected
in our data.

The X-ray spectra were extracted for the source and the background regions
and analyzed  using the fitting software {\sc xspec} \citep{Arnaud1996}.
During the fitting process, the hydrogen column density
$N_{\rm H}$ was fixed according to the stellar reddening, E(B--V)\,=\,0.46,
and using the relation
$N_{\rm H}=6.12\times 10^{21}$\,E(B--V)\,cm$^{-2}$ \citep{Gudennavar2012}.
The emission was modeled using the optically thin, thermal ``mekal''
emission model (\citealt{Liedahl1995}, and references therein)
that assumes collisional ionization equilibrium (CIE). A single temperature
fit to the spectrum did not give a satisfactory fit, and so we adopted a
two-temperature (``2T'') model to characterize the star's X-ray emission.

\begin{figure}
\centering
\includegraphics[height=0.45\textwidth, angle=-90]{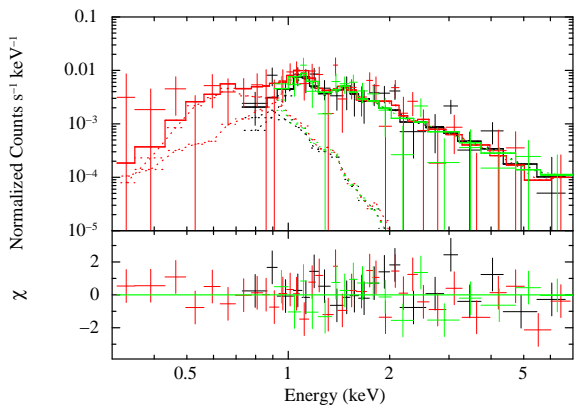}
\caption{
Spectra of CPD\,$-28^{\circ}$\,2561 obtained by the XIS0, XIS1, and XIS3 detectors
on board of {\em Suzaku}, along with a best-fit two-temperature model (see text).
The vertical error bars correspond to 3$\sigma$.
The lower panel shows the quality of the fit in terms of $\chi^2$-fitting statistics.
}
\label{fig:spec}
\end{figure}

Table~\ref{tab:suzaku} summarizes this fit and associated quantities.
The corresponding spectral fit is shown in Fig.\,\ref{fig:spec}.
The observed X-ray flux is given in the band of 0.3--10\,keV as well as
the X-ray luminosity of the source, assuming a distance of 6.1\,kpc. The
X-ray luminosity was corrected for the adopted value of the interstellar
reddening.

Two model components with different temperatures are required to achieve
a satisfactory fit. A soft component with $kT_1 \approx 0.2$\,keV  ($\sim2$\,MK)
is typical of
OB-type stars (e.g.\ \citealt{WaldronCassinelli2007}) including magnetic
stars \citep{Oskinova2011}. Besides this soft emission, we detect
a hard component with $kT_2 \approx 2$\,keV ($\sim24$\,MK).

\begin{table}
\caption{
Spectral parameters derived from the {\em Suzaku} XIS observations of
CPD\,$-28^{\circ}$\,2561 
assuming a two-temperature CIE plasma model ({\it mekal}).  
}
\label{tab:suzaku}
\centering
\begin{tabular}{ll}
\hline
\hline
\multicolumn{1}{c}{Property} &
\multicolumn{1}{c}{Value} \\
\hline
$N_{\rm H}$ [cm$^{-2}$]                            & $2.8 \times 10^{21}$  \\
$f_{\rm X}^{\rm obs}$ [erg\,cm$^{-2}$\,s$^{-1}$ ]  & $1.6 \times 10^{-13}$ \\
$f_{\rm X}^{\rm corr}$ [erg\,cm$^{-2}$\,s$^{-1}$ ] & $5.1 \times 10^{-13}$ \\
$L_{\rm X}$ [erg\,s$^{-1}$]                        & $2.6 \times 10^{33}$  \\
\hline
$\log{L_{\rm X}/L_{\rm bol}}$                      & $-5.7$  \\
\hline
$kT_1$ [keV]                                      & $0.18^{+0.20}_{-0.07}$ \\[1ex]
$EM_1$ [cm$^{-3}$]                                & $8.4^{+4.6}_{-4.6} \times 10^{55}$  \\[1ex]
$kT_2$ [keV]                                      & $2.1^{+0.4}_{-0.3}$  \\[1ex]
$EM_2$ [cm$^{-3}$]                                & $5.5^{+0.8}_{-0.8}  \times 10^{55}$  \\[1ex]
$\chi^2/dof$                                      & 66/70 \\
\hline
$\left< kT \right> [{\rm keV}] \equiv \sum_i kT_i\cdot ({\rm EM}_i/{\rm EM}_{\rm tot})$ & 0.9 \\
\hline
\end{tabular}
\end{table}

The presence of such a hot plasma component with an emission measure that is roughly
half the
emission measure of the cooler spectral component (or about a third of the total 
emission measure), is a special property
that is observed only in a handful of stars.
The average X-ray plasma temperature, $\left< kT \right>$, determined from
spectral fits of CPD\,$-28^{\circ}$\,2561 ($\left< kT \right>\approx0.9$\,keV, 
see Table~\ref{tab:suzaku}) is higher than found even in strongly magnetic B-type stars
\citep{Oskinova2011, Ignace2013}. However, such high temperatures  are deduced
from the analysis of all previously studied X-ray spectra of Of?p-type stars as well
as the O7fp-type star $\theta^1$\,Ori\,C \citep{Schulz2000,naze2014}. The presence of
the hard spectral component can be explained by wind streams from two opposite 
directions of the magnetosphere colliding at the terminal speed $v_\infty$ and 
thus producing strong shocks where the plasma is heated up 
to $24$\,MK \citep{Babel1997, udo2002}.

\section{Conclusions}
\label{sect:concl}

In this work, we used high quality FORS\,2 polarimetric spectra of the Of?p star CPD\,$-28^{\circ}$\,2561 
to carry out magnetic field measurements and to study spectral variability. 
With 20 observations at our disposal, it became now possible to investigate 
the spectral variability of CPD\,$-28^{\circ}$\,2561 over approximately half of the rotation cycle.
The longitudinal 
magnetic field variation can be well represented by the recently reported rotational periodicity
of 73.41\,d.

Using the current mean longitudinal magnetic field measurements and assuming that the star is an 
oblique dipole rotator, we estimate that the polar strength of the surface dipole $B_d$ is
larger than 1.15\,kG.
The comparison of the spectral behaviour of CPD\,$-28^{\circ}$\,2561 with that of another Of?p star, HD\,148937 of
similar spectral type, reveals remarkable differences in the degree of variability between both stars.
Since the reason for the variability in magnetic massive stars is usually referred to rotationally 
modulated emission from a magnetically constrained plasma, it is possible that the low-amplitude spectral
variability of  HD\,148937 is caused by the viewing geometry, i.e. the rotation axis of HD\,148937 must 
be viewed at relatively low inclination, whereas CPD\,$-28^{\circ}$\,2561 is observed at higher inclination 
closer to the stellar equator where plasma emission is optically thick according to the MWC
model by \citet{Babel1997}.
Alternatively, the inclination angle of HD\,148937 could be large while the obliquity between the magnetic field
axis and the rotation angle is small.
 Furthermore, differences in the appearance of the line profiles are detected in the spectra recorded
in different years, indicating that profiles are not strictly repeatable due to either uncertainties in 
the rotation period or due to intrinsic variability caused by changes in the 
amount or distribution of emitting material with time in the magnetosphere of CPD\,$-28^{\circ}$\,2561.
Additional spectra and  spectropolarimetric observations over the whole rotation cycle are needed
to better constrain the magnetic field model of this star.

Our X-ray observations obtained with the {\em Suzaku X-ray Observatory} agree well with the supposition of 
the presence of the magnetically confined stellar wind as we find evidence for a hard component 
with $kT_2 \approx2$\,keV. This component is much harder than typical for a single non-magnetic OB stars. 
Unfortunately, as our observations had the total duration of about 7.5\,h, no conclusion can be drawn
from these data about the rotation modulation of the X-ray flux.
One important task for future X-ray  studies is to obtain detailed information on the variability of X-rays and to 
search for the periodicity usually expected in massive stars with the X-ray emitting confined winds. 

\section*{Acknowledgments}
We thank the referee John Landstreet for useful comments to the original manuscript.
Based on observations collected with the {\em Suzaku X-ray Observatory}, data 
obtained with ESO telescopes at the La Silla Paranal Observatory under programme
ID 092.D-0209(A), and data obtained from the ESO Science Archive Facility under request number 
MSCHOELLER51303. 
A.F.K thanks for support by the Saint-Petersburg University project 6.38.18.2014.

\end{document}